\documentclass[prl,amsmath,amssymb,twocolumn,superscriptaddress]{revtex4-2}
\usepackage{amsmath}
\usepackage{amssymb}
\usepackage{amstext}
\usepackage{amsfonts}
\usepackage{amsxtra}
\usepackage{bm}
\usepackage[usenames]{color}
\usepackage{eufrak}
\usepackage{grffile}
\usepackage[hidelinks]{hyperref}
\usepackage{epstopdf}
\usepackage{soul}
\usepackage[normalem]{ulem} 
\usepackage{cleveref}

\newcommand{\dx}{\text{d}^3\textbf{x}\,}
\newcommand{\edd}{\varepsilon_\text{dd}}

\begin{document}

\title{Two-dimensional supersolid formation in dipolar condensates}

 \author{T. Bland}
  \affiliation{
      Institut f\"{u}r Quantenoptik und Quanteninformation, \"Osterreichische Akademie der Wissenschaften, Innsbruck, Austria
  }
 
 \author{E. Poli}
  \affiliation{
      Institut f\"{u}r Experimentalphysik, Universit\"{a}t Innsbruck, Austria
  }
 
 \author{C. Politi}
  \affiliation{
      Institut f\"{u}r Quantenoptik und Quanteninformation, \"Osterreichische Akademie der Wissenschaften, Innsbruck, Austria
  }
  \affiliation{
      Institut f\"{u}r Experimentalphysik, Universit\"{a}t Innsbruck, Austria
  }
 
 \author{L. Klaus}
  \affiliation{
      Institut f\"{u}r Quantenoptik und Quanteninformation, \"Osterreichische Akademie der Wissenschaften, Innsbruck, Austria
  }
  \affiliation{
      Institut f\"{u}r Experimentalphysik, Universit\"{a}t Innsbruck, Austria
  }

 \author{M. A. Norcia}
  \affiliation{
      Institut f\"{u}r Quantenoptik und Quanteninformation, \"Osterreichische Akademie der Wissenschaften, Innsbruck, Austria
  }

 \author{F. Ferlaino}
  \affiliation{
      Institut f\"{u}r Quantenoptik und Quanteninformation, \"Osterreichische Akademie der Wissenschaften, Innsbruck, Austria
  }
  \affiliation{
      Institut f\"{u}r Experimentalphysik, Universit\"{a}t Innsbruck, Austria
  }

 \author{L. Santos}
  \affiliation{
      Institut f\"{u}r Theoretische Physik, Leibniz Universit\"{a}t Hannover, Germany
  }

 \author{R. N. Bisset}
  \affiliation{
      Institut f\"{u}r Experimentalphysik, Universit\"{a}t Innsbruck, Austria
  }

\begin{abstract}

Dipolar condensates have recently been coaxed to form the long-sought supersolid phase.
While one-dimensional (1D) supersolids may be prepared by triggering a roton instability, we find that such a procedure in two dimensions (2D) leads to a loss of both global phase coherence and crystalline order.
Unlike in 1D, the 2D roton modes have little in common with the supersolid configuration.
We develop a finite-temperature stochastic Gross-Pitaevskii theory that includes beyond-meanfield effects to explore the formation process in 2D, and find that evaporative cooling directly into the supersolid phase--hence bypassing the first-order roton instability--can produce a robust supersolid in a circular trap. Importantly, the resulting supersolid is stable at the final nonzero temperature.
We then experimentally produce a 2D supersolid in a near-circular trap through such an evaporative procedure. Our work provides insight into the process of supersolid formation in 2D, and defines a realistic path to the formation of large two-dimensional supersolid arrays.


\end{abstract}
\date{\today}
\maketitle

The supersolid phase was predicted to simultaneously exhibit crystalline order and superfluidity \cite{gross1957unified,andreev1969quantum,thouless1969flow,chester1970speculations,leggett1970can,boninsegni2012colloquium}. While it remains elusive in helium, recent developments in ultracold quantum gases have finally made supersolidity a reality, providing an excellent platform for the control and observation of these states.
Important early advances were made in systems with spin-orbit coupling \cite{li2017stripe,Bersano2019} and cavity-mediated interactions \cite{leonard2017supersolid}, where supersolid properties were observed in rigid crystal configurations.
Bose-Einstein condensates (BECs) with dipole-dipole interactions have now been observed in a supersolid state with deformable crystals \cite{Tanzi2019,Bottcher2019,Chomaz2019,norcia2021two}, with their lattices genuinely arising from the atom-atom interactions \cite{natale2019excitation,tanzi2019supersolid,guo2019low}.

In the first dipolar supersolid experiments translational symmetry was broken only along one axis, giving rise to a one-dimensional (1D) density wave, commonly referred to as a 1D droplet array \cite{Tanzi2019,Bottcher2019,Chomaz2019}. A more recent experiment has created the first states with two-dimensional (2D) supersolidity in elongated traps of variable aspect ratio \cite{norcia2021two}.
This opens the door to study vortices and persistent currents \cite{gallemi2020quantized,roccuzzo2020rotating,tengstrand2021persistent,ancilotto2021vortex}, as well as exotic ground state phases predicted for large atom numbers \cite{Baillie2018,zhang2019supersolidity,zhang2021phases,hertkorn2021pattern}.

It is still an open question whether 2D arrays provide as favorable conditions for supersolidity as 1D arrays do.
In 1D, following an interaction quench from an unmodulated to modulated BEC, the density pattern induced by a roton instability \cite{santos2003roton,Giovanazzi2004,Petter2019,Chomaz2018a,natale2019excitation} can smoothly connect with the final supersolid array \cite{Tanzi2019,Bottcher2019,Chomaz2019}. This transition hence has a weakly first-order character, or is even continuous \cite{blakie2020supersolidity,biagioni2021dimensional}, and such quenches through the transition cause only small excitations of the resulting supersolid \cite{Bottcher2019,Tanzi2019,Chomaz2019}.
While it has been predicted that a similar procedure may lead to coherence between three droplets in a triangular configuration \cite{hertkorn2021supersolidity},
earlier work with nondipolar superfluids suggests that such symmetry breaking quenches may be unfavorable for supersolid formation in 2D and 3D \cite{Macri2013,Henkel2010}.

An alternative method exists to experimentally produce dipolar supersolids. Instead of quenching the interactions to trigger a roton instability,
it is possible to cool a thermal sample directly into the supersolid state using evaporative cooling techniques \cite{Chomaz2019,Sohmen2021}.
Crucially, this is the only known method for producing 2D supersolids to date \cite{norcia2021two}.
While a dynamic interaction quench may be described by the extended Gross-Pitaevskii equation (eGPE) \cite{Wachtler2016a,Bisset2016,Ferrier-Barbut2016a,Chomaz2016}, we are not aware of any available theory to model the required evaporation process.
From a theoretical perspective much remains unknown about evaporative supersolid formation. Is it a general feature that the droplets form before global phase coherence develops, as reported in \cite{Sohmen2021}? Under what conditions do defects persist? Such answers will be paramount in the quest for ever-larger 2D supersolids, as well as for the observation of vortices embedded within them.


\begin{figure*}[!t]
        \includegraphics[width=0.7\textwidth]{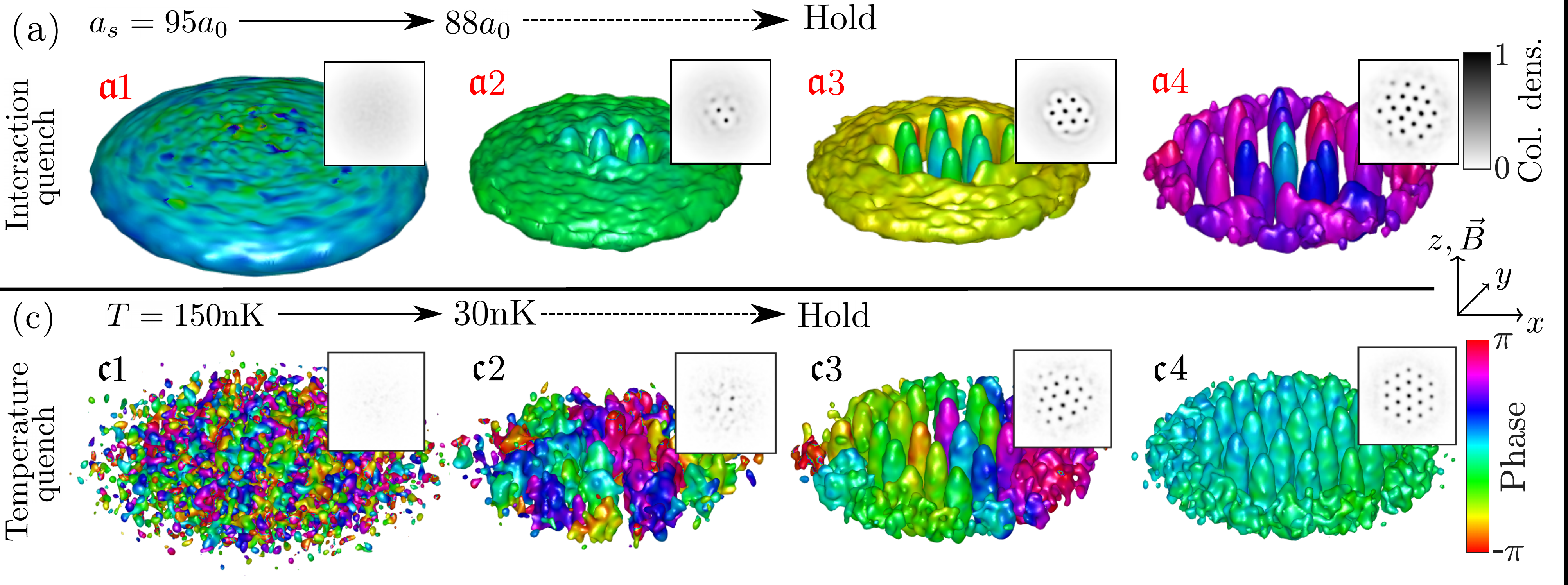}
       \includegraphics[width=0.28\textwidth]{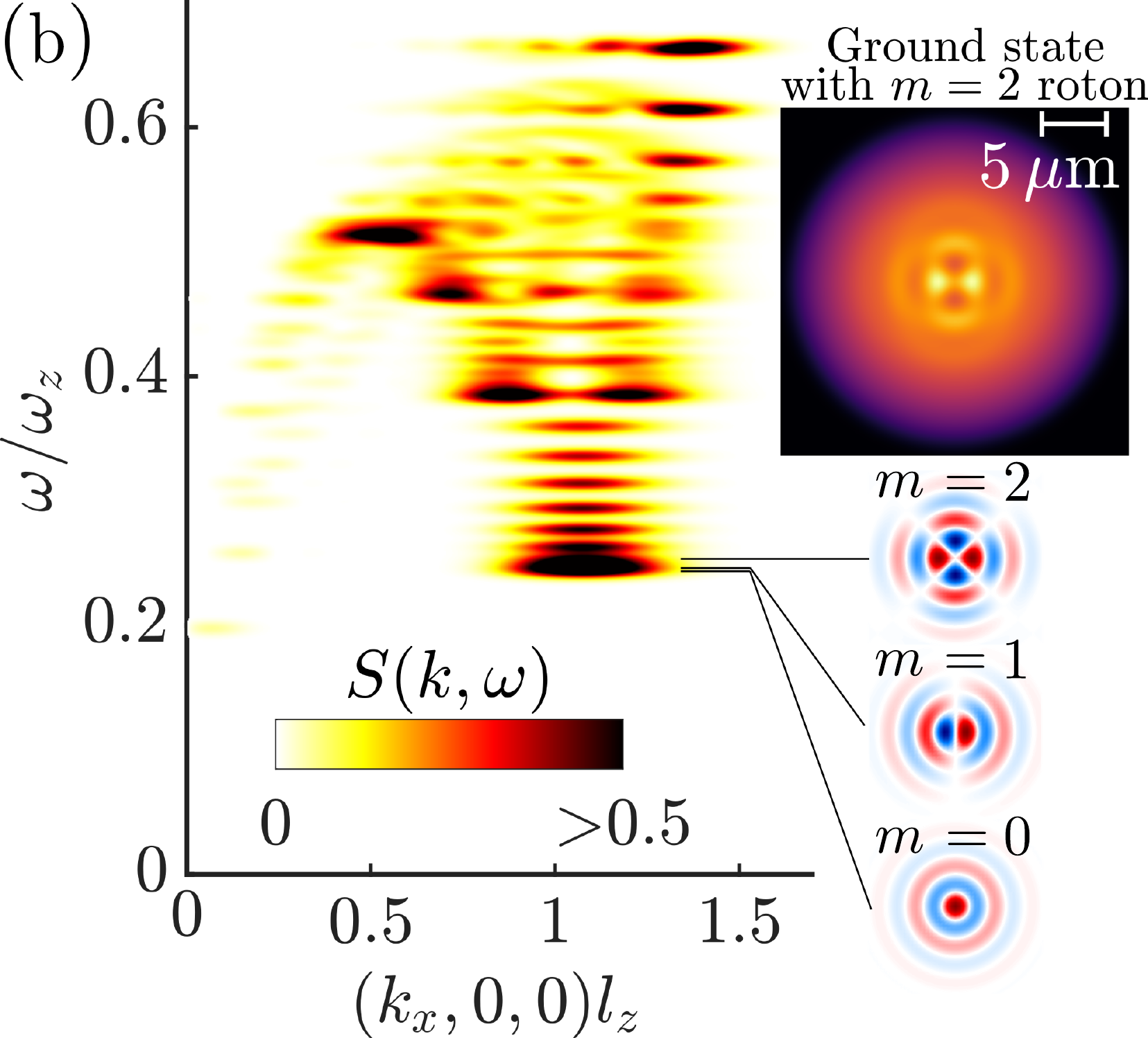}
\caption {
(a) Crystal preparation from interaction quench, evolved with the eGPE, for $N \approx 2.1\times10^5$ Dy atoms [quench (i)]. Isosurfaces are at 5\% max density, with color indicating phase. Insets: $z$ column densities normalized to max value from entire simulation. (b) Dynamic structure factor for unmodulated BEC ($a_s=92a_0$) in energy-momentum space, normalized to peak value. The lowest energy roton modes are indicated, and the ground state with an $m=2$ roton mode added is shown, revealing the localized nature of the rotons. Parameters are otherwise the same as in (a).
(c) Crystal preparation from temperature quench (evaporative cooling) evolved with the SeGPE [quench (ii)].
The temperature decreases as the chemical potential and condensate number rise, with scattering length fixed at $a_s=88a_0$.
For all subplots $f_{x,y,z}= (33,33,167)$\,Hz, and $l_z=\sqrt{\hbar/m\omega_z}\,$.}
	 \label{fig:1} 
\end{figure*}

In this work, we explore the formation of large 2D supersolids in circular-shaped traps.
We develop a finite temperature Stochastic eGPE (SeGPE) theory to model the entire evaporative cooling process. Importantly, our theory includes the beyond-meanfield quantum fluctuations responsible for stabilizing the individual droplets.

We compare the evaporative cooling formation dynamics with those resulting from an interaction quench, finding striking differences between the two protocols.
Following an interaction quench, the 2D crystal grows nonlinearly with the droplets developing sequentially, producing configurations that are unrelated to any roton mode combination of the original unmodulated BEC.
The resulting crystal is substantially excited and lacks global phase coherence.
Alternatively, by directly cooling into the supersolid regime our SeGPE theory predicts the formation of large 2D supersolids in circular traps, with global phase coherence that remains robust at finite temperature.
To benchmark our theory--as well as to test the direct cooling protocol for pancake-shaped trapping geometries--we perform experiments and observe a 7-droplet hexagonal supersolid in a near-circular trap.



{\it Formalism}.---We are interested in ultracold, dipolar Bose gases harmonically confined in 3D with trapping frequencies $\omega_{x,y,z}=2\pi f_{x,y,z}$.
Two-body contact interactions and the long-ranged, anisotropic dipole-dipole interactions are well-described by a pseudo-potential,
$U(\textbf{r}) = \frac{4\pi\hbar^2a_{\rm s}}{m}\delta(\textbf{r})+\frac{3\hbar^2a_\text{dd}}{m}\frac{1-3\cos^2\theta}{r^3}$, with $a_{\rm s}$ being the s-wave scattering length and $a_\text{dd}=\mu_0\mu_m^2m/12\pi\hbar^2$ is the dipole length, with magnetic moment $\mu_m$, and $\theta$ is the angle between the polarization axis ($z$ axis) and the vector joining two particles.
The ratio $\epsilon_\text{dd} = a_\text{dd}/a_\text{s}$ (for $a_\text{s}>0$) is convenient to keep in mind since for $\epsilon_\text{dd}\leq 1$ the ground state will be an unmodulated BEC, whereas for the dipole-dominated regime $\epsilon_\text{dd}>1$ the unmodulated BEC may become unstable \cite{Lahaye_RepProgPhys_2009}.
Here, we always consider $^{164}$Dy, with $a_\text{dd}=131a_0$.
The eGPE has been described elsewhere \cite{Wachtler2016a,Bisset2016,Ferrier-Barbut2016a,Chomaz2016}, and its details have been deferred to the supplementary material \cite{supp}.

We phenomenologically introduce a finite-temperature simple growth SeGPE theory \footnote{This method has been applied previously in a quasi-two-dimensional setting without quantum fluctuations \cite{Linscott2014a}.}. This describes the `classical' field, $\Psi({\mathbf r},t)$, of all highly-populated modes up to an energy cut-off. The dynamics are governed by~\cite{Blakie2008}:
\begin{align}
    i\hbar\frac{\partial\Psi}{\partial t} = \hat{\mathcal{P}}\Big\{ \Big(1- i\gamma\Big)(\mathcal{L}[\Psi]-\mu)\Psi + \eta\Big\}\,.
    \label{eqn:spgpe}
\end{align}
Here, $\mathcal{L}$ is the eGPE operator defined in \cite{supp}, and $\gamma$ describes the coupling of the classical field modes to the high-lying modes.
We find that ${\gamma=7.5\times10^{-3}}$ gives good agreement to the condensate number growth rate of a recent experiment under comparable conditions \cite{Sohmen2021} (see also \cite{supp}). The dynamical noise term $\eta$, subject to noise correlations given by $\langle \eta^{*}(\textbf{r},t)\eta(\textbf{r}',t') \rangle = 2\hbar\gamma k_{\textrm{B}}T\delta(t-t')\delta(\textbf{r}-\textbf{r}')$, means that each simulation run is unique.
Finally, $\hat{\mathcal{P}}$ is a projector which constrains the dynamics of the system up to energy cutoff $\epsilon_\mathrm{cut} (\mu) =2\mu$--consistent with previous treatments \cite{Rooney2013,mcdonald2020dynamics}--where we use the final $\mu$ after evaporative cooling.

{\it Supersolid formation simulations}.---With these two theories in hand, we perform two kinds of dynamic quench simulations in a pancake-shaped trap, where in both cases the ground state for the final parameters would be a 19-droplet supersolid:

(i) An interaction quench from an unmodulated BEC to the supersolid regime using the eGPE [Fig.~\ref{fig:1}(a)].
Noise is first added to the BEC ground state \footnote{Our initial state is $\psi(\textbf{r},0) = \psi_0(\textbf{r}) + \sum_n' \alpha_n \phi_n(\textbf{r})$, where $\phi_n$ are the single-particle states, $\alpha_n$ is a complex Gaussian random variable with $\langle|\alpha_n|^2\rangle = (e^{\epsilon_n/k_BT}-1)^{-1}+\frac12$ and the sum is restricted to modes with $\epsilon_n\le2k_BT$, with $T=30$nK.},
and this is evolved for a $20$ms equilibration time before the interaction strength is linearly ramped over the next $30$ms from $a_s = 95a_0$ to $a_s = 88a_0$--crossing the roton phase transition to the supersolid regime--and then held constant again for the remainder of the simulation.

(ii) A temperature quench from a thermal cloud to the supersolid phase using the SeGPE [Fig.~\ref{fig:1}(c)].
Each simulation begins with a 200ms equilibration time at fixed high temperature $T=150$nK to generate a thermal cloud.
To simulate the evaporative cooling process, the chemical potential and temperature are then linearly ramped over 100ms, from $(\mu,\,T)$=$(-12.64\hbar\omega_z,\,150$nK) to $(12.64\hbar\omega_z,\,30$nK), mimicking the growing condensate number observed in experiments \cite{Weiler08a,liu2018dynamical}, whilst the scattering length is always held fixed at $a_s = 88a_0$.

Focusing first on the interaction quench, the density isosurfaces in Fig.~\ref{fig:1}(a) represent snapshots at various times for a single simulation run,
revealing intriguing formation dynamics.
Initial droplets are seeded through unstable roton modes, but staggered droplet formation reveals a process of {\it nonlinear crystal growth}, as highlighted by the column densities shown as insets to Fig.~\ref{fig:1}(a).
In Fig.~\ref{fig:1}(a2), two central droplets have already attained their final peak density, while a secondary ring of droplets is only just beginning to form.
Then in Fig.~\ref{fig:1}(a3), eight droplets have fully matured and the process continues radially outwards until a 19 droplet crystal is approximately attained. Similar droplet formation dynamics have been predicted in optical media \cite{maucher2016self}.

The colors on the density isosurfaces in Figs.~\ref{fig:1}(a) represent the wavefunction phase. The colorscale is re-centered in each subplot and an ideal phase coherent solution would have a uniform color everywhere. Importantly, the crystal growth process disrupts the global phase coherence, as evidenced by the various colors in Fig.~\ref{fig:1}(a4), leaving an excited crystal in which some outer droplets dissolve and reemerge from the halo.
Note that the situation does not qualitatively change for reduced initial noise or gentler interaction ramps, suggesting that the strong excitations result from a first-order character of the roton instability in 2D.


We explain the interaction quench dynamics by calculating the elementary excitations of the unmodulated BEC close to the roton instability, i.e., for $a_{\text s} = 92 a_0$.
These results are displayed in Fig.~\ref{fig:1}(b) as the dynamic structure factor $S(\textbf{k},\omega)$, which predicts the system response to perturbations of momentum $\hbar \textbf{k}$ and energy $\hbar\omega$ \cite{Zambelli2000,blakie2002theory,Blakie2012a,Chomaz2018a} (also see \cite{supp}).
A roton minimum can be seen at $k_xl_z\approx1.1$, and we plot the lowest roton modes corresponding to $m=0,1,2$, with $m$ being the angular quantum number in the $z$ direction \cite{ronen2007radial}.
On the top-right is the density obtained by adding an $m=2$ roton mode to the BEC wavefunction, revealing how rotons are confined to high-density regions \cite{JonaLasinio2013,bisset2013roton}.
This reveals a qualitative difference between the 1D and 2D situations since, from a simple geometric standpoint, in 2D the high-density region inherently encompasses a smaller proportion of the total atom number.
Thus, the roton-induced droplet number is only a small fraction of the final droplet number, meaning the droplets appear sequentially for 2D.

\begin{figure}[!t]
    \includegraphics[width=0.5\textwidth]{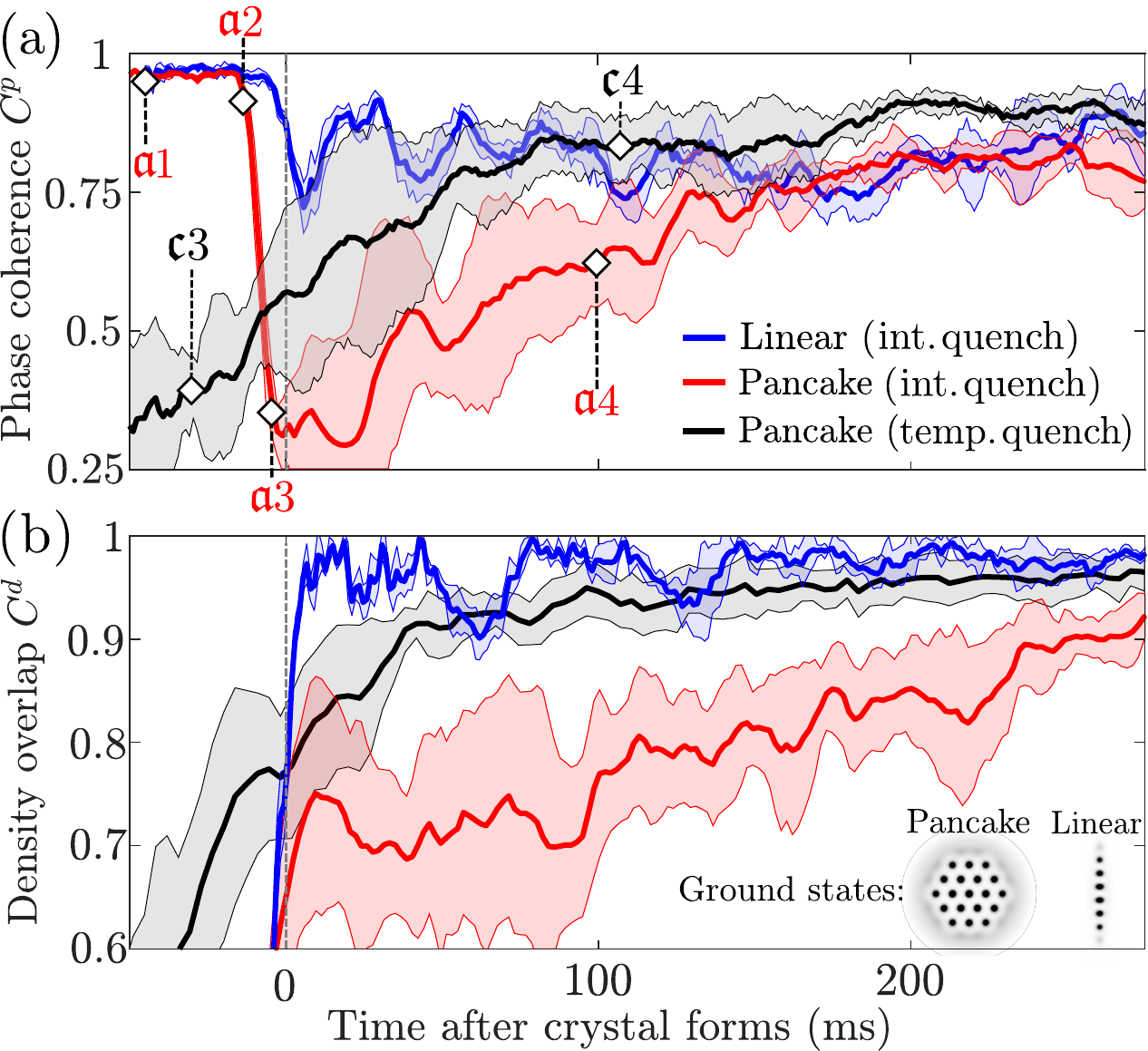}
\caption{Supersolid quality. (a) Phase coherence of droplets $C^p$ over time for interaction quenches [quench (i)] into linear chain (blue) and pancake crystal (red), and temperature quenches [quench (ii)] into the pancake crystal (black). Diamonds link to example frames in Figs.~\ref{fig:1}(a,c). Each curve is averaged over 3-5 runs with error band marking one standard deviation.
Time $t=0$ corresponds to when the crystals first fully mature.  (b) Density overlap $C^d$ between the time-dependent and ground state densities. Parameters same as Fig.~\ref{fig:1}, but for linear chain $f_{x,y,z}= (33,110,167)$\,Hz and $N=82\times10^3$.}
\label{fig:2}
\end{figure}

Another qualitative difference between 1D and 2D is a kind of frustration.
First, note that our target supersolid ground state for the final quench parameters is a 19 droplet crystal, with a central droplet [see bottom of Fig.~\ref{fig:2}(b)].
Only an $m=0$ roton mode [see Fig.~\ref{fig:1}(b)] could directly trigger the formation of a central droplet, but then only concentric rings could form further out.
Thus, unlike for 1D, no single roton mode can smoothly connect the unmodulated BEC to the 2D supersolid ground state.

Next, we analyze the finite-temperature quench results. Figure \ref{fig:1}(c) shows snapshots of the condensate growth,
demonstrating that both the crystal structure and the global phase coherence--evidenced by the uniform color in Fig.~\ref{fig:1}(c4)--develop soon after the quench.
Note that time scales will be quantified shortly.
It is also an important result in itself that we predict such a large 2D supersolid to be stable against thermal fluctuations (recall that $T_{\text{final}}=30$nK).
As they form, each droplet individually has a uniform phase that may be different from that of its neighbors, sometimes creating vortex pairs between droplets of different phase. In this scenario, the droplets do not form as a result of a roton instability, and the partial phase coherence continues to improve after the crystal has formed, consistent with earlier observations \cite{Sohmen2021}.
Occasionally, long-lived isolated vortices remain near the centre of the supersolid.  Simulation videos are provided in supplemental material \cite{supp}.

{\it Supersolid quality}.---We seek to quantify the resulting supersolid quality for both quench protocols. We start by analyzing the phase excitations, taking the phase coherence $C^p$ with a similar measure presented in Ref.~\cite{Tanzi2019}.
A value of $C^p = 1~(0)$ implies global phase coherence (incoherence) \footnote{We measure phase coherence as $C^p = 1 - \frac{2}{\pi}\int_\mathcal{C}\text{d}x\text{d}y\,|\psi(x,y)|^2|\theta(x,y)-\beta|/\int_\mathcal{C}\text{d}x\text{d}y\,|\psi(x,y)|^2$, where $\theta(x,y)$ is the phase of $\psi(x,y)$ in the $z=0$ plane, and $\beta$ is a fitting parameter to maximize $C^p$ at each time. The integration region $\mathcal{C}$ encompasses the droplets.}.
In Fig.~\ref{fig:2}(a), we plot this quantity for interaction quenches into the pancake supersolid regime (red) and linear supersolid regime (blue), and temperature quenches into the pancake supersolid (black).
The time $t=0$ indicates when the droplet number has approximately stabilized, and the crystal has first matured \footnote{Relative to the crystal maturation time, each quench ramp ended at 5ms, -13ms, and -90ms for the linear interaction quench, pancake interaction quench, and pancake temperature quench, respectively. Note that evaporative cooling continues after the \{$\mu$, $T$\} quench ends, taking approximately a further 100ms until the classical field modes have equilibrated with the high-energy modes \cite{supp}.}.
For the linear chain, the system remains coherent (high $C^p\approx 0.8$), indicating a stable supersolid.
However, quenching into the pancake geometry is qualitatively different, with strong incoherence ($C^p\approx 0.3$) soon after crystal formation, recovering a high value at around 150ms after the crystal forms. During evaporative cooling, the global phase coherence is predicted by the high value of $C^p\approx 0.8$ around 50ms after the crystal forms, with qualitatively similar values to the interaction quench simulations for the linear supersolid case.

We quantify the quality of the supersolid crystal by measuring the density overlap $C^d$ between the ground state target solution and the time dependent wavefunction \footnote{The density overlap is given by $C^d = \int \text{d}^3\textbf{x}\, n(\textbf{x} - \textbf{x}_0,\phi_0,t)n_\text{GS} / \int \text{d}^3\textbf{x}\, n_\text{GS}^2$, with the ground state density and time-dependent density, respectively, normalized as $\int \text{d}^3\textbf{x}\, n_\text{GS} = \int \text{d}^3\textbf{x}\, n(\textbf{x} - \textbf{x}_0,\phi_0,t) = 1$. The optimization parameters $\textbf{x}_0$ and $\phi_0$ are translations and a rotation, respectively, applied to the wavefunction to maximize $C^d$.}. We find the maximal value of $C^d$ after applying translations and rotations to the state, noting that perfect overlap would give $C^d=1$. In Fig.~\ref{fig:2}(b) this quantity is presented for the two geometries, with the ground state solutions shown as insets.
For the linear chain, once the droplets have formed the density overlap rapidly attains $C^d>0.9$ and remains there, consistent with the interaction quenched state being close to the ground state supersolid. However, the pancake case shows weak overlap after the droplets are formed, which only recovers slowly, after around 300ms, to values comparable with the linear chain.
Primarily, this is due to the sensitivity of droplet positions of $C^d$, and indicates that there are many excited supersolid modes present after the droplets form \cite{supp}.
Direct evaporative cooling for the pancake case, however, shows that after the droplets have formed they rapidly settle into the expected crystal pattern ($C^d\approx0.95$).

Finally, it is important to note that for the pancake interaction quench, while the phase coherence is restored by around $t=150$ms after the droplets are formed, the crystal remains highly excited until around 300ms.
On these timescales three-body losses become significant, and it is unlikely that a large supersolid would be observed. In contrast, direct evaporative cooling may lead to a robust supersolid within around 50ms of the crystal first appearing, a timescale that we find to be weakly dependent on the value of $\gamma$ \cite{supp}.

\begin{figure}[!t]
    \centering
\includegraphics[width=3.0in,angle =-90]{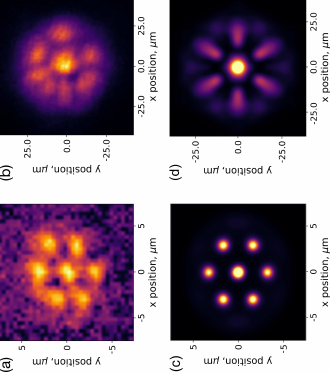}
\caption {Experimental realization of seven droplet hexagon state.  (a) Exemplary in-situ image of density profile.  (b) Image after 36~ms time-of-flight (TOF) expansion, averaged over 68 trials of the experiment.  Hexagonal modulation structure is clearly present in the averaged image. Note the rotation of the hexagon between in-situ and TOF images. (c,d) Corresponding simulations for same trap, and with $a_s=90a_0$ and $\approx 4.4\times10^4$ atoms within the droplets.}
	 \label{fig:3} 
\end{figure}

{\it Experimental observation}.---While experiments have evaporatively cooled directly into the supersolid phase for linear and elongated 2D configurations \cite{Chomaz2019,Sohmen2021,norcia2021two},
this could prove an optimal method in circular traps for avoiding the excitations associated with crossing the roton instability.
We confirm this by producing a 7 droplet hexagon supersolid in a near-circular trap, as shown in Fig.~\ref{fig:3}. The experimental apparatus and procedure is similar to that described previously \cite{norcia2021two}, but new modifications in the optical dipole trap setup have enabled us to tune between anisotropic and round traps. The current optical trap consists of three 1064~nm wavelength trapping beams, each propagating in the plane perpendicular to gravity.  Two of the beams, which cross perpendicularly, have approximately 60~$\mu$m waists, and define the horizontal trapping frequencies. The third, crossing at a roughly 45 degree angle from the others, has a waist of approximately 18~$\mu$m and is rapidly scanned to create a time-averaged light sheet that defines the vertical confinement.

In a harmonic trap with frequencies $f_{x,y,z}  = (47(1), 43(1), 133(5))$Hz, we observe in-trap a 7-droplet state consisting of a hexagon with a central droplet, with a condensate atom number of $N\sim4\times10^4$ [Fig.~\ref{fig:3}(a)]. To confirm that this state is phase-coherent, we release the atoms from the trap and image the interference pattern after 36~ms time-of-flight [Fig.~\ref{fig:3}(b)].  The presence of clear modulation in the interference pattern averaged over 68 runs of the experiment indicates a well-defined and reproducible relative phase between the droplets, and is consistent with our expectations for a phase-coherent state undergoing expansion [Fig.~\ref{fig:3}(d)], obtained through 3D dynamic simulations starting from the eGPE ground state [Fig.~\ref{fig:3}(c)]. Even rounder traps are possible, but the slight anisotropy orients the state, helping to observe the reproducible interference pattern.


{\it Summary}.---We have theoretically explored the formation of large 2D supersolids using both an interaction quench from an unmodulated BEC, and a temperature quench from a thermal cloud. For the latter, we developed a finite-temperature stochastic Gross-Pitaevskii theory that can simulate evaporative cooling directly into the supersolid regime.
Our simulations predict that a temperature quench provides a robust path for creating 2D supersolids in circular traps, and we confirm this experimentally by using this method to create a reproducible hexagonal 7-droplet supersolid.

In contrast, the interaction quench results in highly-excited crystals that lack global phase coherence in the period following their formation.
Interestingly, droplets appear sequentially rather than simultaneously,
with the final crystal structure being unrelated to the roton modes that seeded the instability.
This is in contrast to the situation for 1D arrays, where an interaction quench through a roton instability can smoothly connect an unmodulated BEC to the supersolid ground state.

Our finite-temperature theory is broadly applicable for future studies on topics such as formation dynamics, supersolid vortices, improved quench protocols to produce large 2D supersolids, thermal resilience, as well as dipolar droplets in general.


\begin{acknowledgements}
{\it Acknowledgements}.---We thank Manfred Mark and the Innsbruck Erbium team for valuable discussions, and thank P\'eter Juh\'asz for carefully reading the manuscript. We acknowledge R.~M.~W.~van Bijnen for developing the code for our eGPE and BdG simulations. Part of the computational results presented have been achieved using the HPC infrastructure LEO of the University of Innsbruck. The experimental team is financially supported through an ERC Consolidator Grant (RARE, No.~681432), an NFRI grant (MIRARE, No.~OAW0600) of the Austrian Academy of Science, the QuantERA grant MAQS by the Austrian Science Fund FWF No I4391-N. L.S. and F.F.~acknowledge the DFG/FWF (FOR 2247/I4317-N36) and a joint-project Grant from the FWF (I4426, RSF/Russland 2019). L.S.~thanks the funding by the Deutsche Forschungsgemeinschaft (DFG, German Research Foundation) under Germany’s Excellence Strategy--EXC-2123 QuantumFrontiers--390837967. M.A.N.~has
received funding as an ESQ Postdoctoral Fellow from the European Union’s Horizon 2020 research and innovation programme under the Marie Sklodowska-Curie grant agreement No.~801110 and the Austrian Federal Ministry of Education, Science and Research (BMBWF). We also acknowledge the Innsbruck Laser Core Facility, financed
by the Austrian Federal Ministry of Science, Research and Economy.
\end{acknowledgements}


%

\clearpage
\appendix
\onecolumngrid
\begin{center}
    {\bf\large Supplemental materials: Two-dimensional supersolid formation in dipolar condensates}\\\vspace{0.3cm}
    {\normalsize T. Bland, E. Poli, C. Politi, L. Klaus, M. A. Norcia, F. Ferlaino, L. Santos, and R. N. Bisset}
\end{center}
\hspace{3cm}
\twocolumngrid
\makeatletter
\renewcommand{\theequation}{S\arabic{equation}}
\renewcommand{\thefigure}{S\arabic{figure}}
\setcounter{equation}{0}
\setcounter{figure}{0}

\normalsize

\renewcommand{\theequation}{S\arabic{equation}}
\renewcommand{\thefigure}{S\arabic{figure}}

\section{Formalism}

We utilize dynamic and ground state calculations of the extended Gross-Pitaevskii equation (eGPE), given by $i\hbar \psi_t = \mathcal{L}[\psi]\psi$, where the eGPE operator is \cite{Wachtler2016a,Bisset2016,Ferrier-Barbut2016a,Chomaz2016}
\begin{align}
    &\mathcal{L}[\psi]=-\frac{\hbar^2\nabla^2}{2m}+\frac12m\left(\omega_x^2x^2+\omega_y^2y^2+\omega_z^2z^2\right) \label{eqn:HGP} \\
    &\quad+ \int\text{d}^3\textbf{x}'\, U(\textbf{x}-\textbf{x}')|\psi(\textbf{x}',t)|^2  +\gamma_\text{QF}|\psi(\textbf{x},t)|^3\,, \notag
\end{align}
$m$ is the mass and $\omega_{x,y,z}=2\pi f_{x,y,z}$ are the external trapping frequencies. Two-body contact interactions and the long-ranged, anisotropic dipole-dipole interactions are described by the pseudo-potential
\begin{align}
    U(\textbf{r}) = \frac{4\pi\hbar^2a_{\rm s}}{m}\delta(\textbf{r})+\frac{3\hbar^2a_\text{dd}}{m}\frac{1-3\cos^2\theta}{r^3}\,,
\end{align}
respectively, with $\theta$ being the angle between the polarization axis ($z$ axis) and the vector joining two particles.
This is characterized by s-wave scattering length $a_{\rm s}$ and dipole length ${a_\text{dd}=\mu_0\mu_m^2m/12\pi\hbar^2}$, with magnetic moment $\mu_m$.
To find the ground state we employ a conjugate-gradients technique minimizing the corresponding energy functional \cite{Ronen2006a}.
The last term appearing in Eq.~\eqref{eqn:HGP} represents quantum fluctuations in the form of a dipolar Lee-Huang-Yang correction \cite{Lima2011a}, $\gamma_\text{QF}=\frac{128\hbar^2}{3m}\sqrt{\pi a_s^5}\,\text{Re}\left\{ \mathcal{Q}_5(\edd) \right\}$, where $\mathcal{Q}_5(\edd)=\int_0^1 \text{d}u\,(1-\edd+3u^2\edd)^{5/2}$ is the auxiliary function, which can be solved analytically, and $\edd = a_{\rm dd}/a_{\rm s}$.

Primarily, we use the eGPE to simulate the formation dynamics of a supersolid through an interaction quench of the scattering length, constructing an initial state by adding non-interacting noise to an unmodulated BEC ground state (far from the roton instability). Thus, our initial state is $\psi(\textbf{r},0) = \psi_0(\textbf{r}) + \sum_n' \alpha_n \phi_n(\textbf{r})$, where $\phi_n$ are the single-particle states, $\alpha_n$ is a complex Gaussian random variable with $\langle|\alpha_n|^2\rangle = (e^{\epsilon_n/k_BT}-1)^{-1}+\frac12$ with temperature $T$ and the sum is restricted to modes with $\epsilon_n\le2k_BT$. On average, this adds about 1000 atoms when $T=30$nK.

\section{Choice of $\gamma$ for Stochastic extended GPE theory (Eq.~1 of main text)}

\begin{figure}[!t]
    \centering
    \includegraphics[width = 0.49\textwidth]{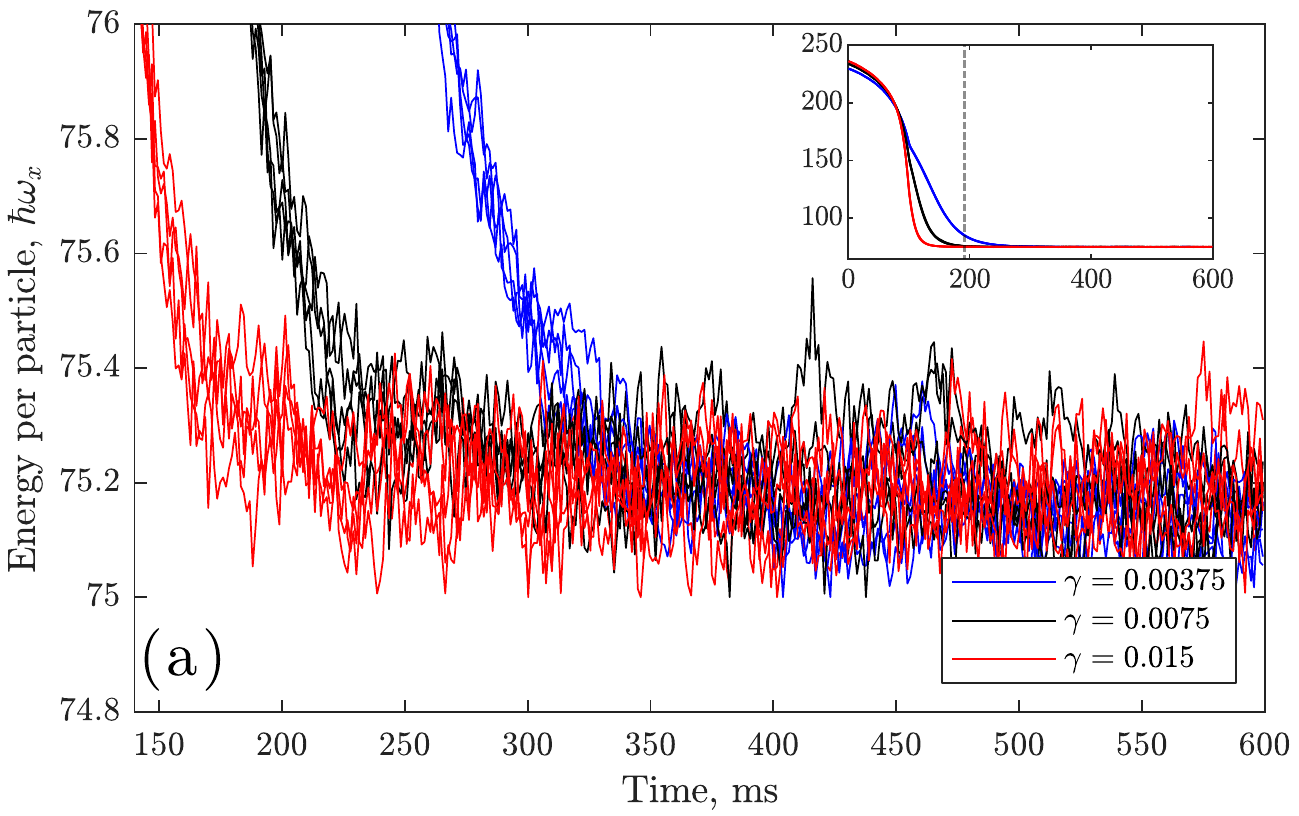}
    \includegraphics[width = 0.49\textwidth]{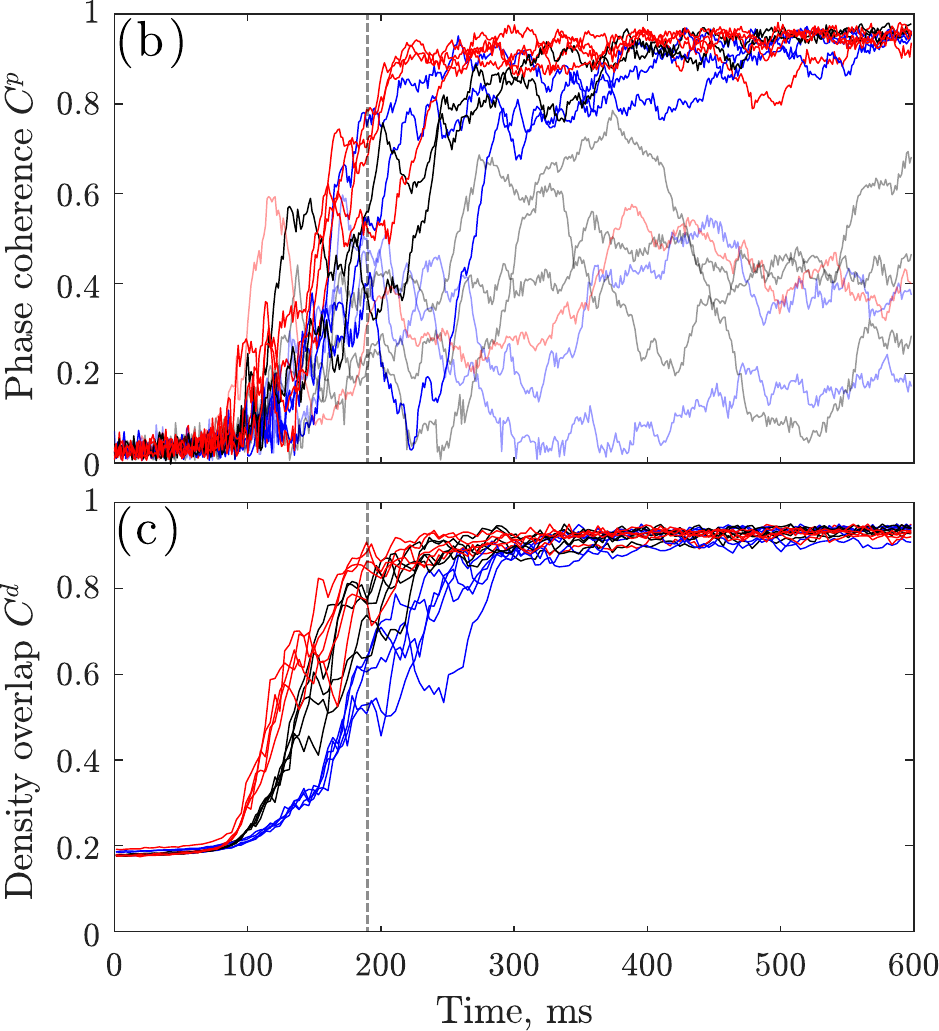}
    \caption{Effect of varying $\gamma$. Five simulations are shown with the same color for each $\gamma$. (a) Energy per particle versus time. The average final value is independent of $\gamma$, as expected. Inset shows the same data over a longer time range, where $t=0$ corresponds to the beginning of the 100ms temperature quench. (b) Phase coherence for all data. Curves with a lower opacity correspond to simulations with isolated vortices. (c) Density overlap with the ground state 19 droplet supersolid. In all plots the dashed vertical line indicates $t=0$ in Fig.~2 of the main text.}
    \label{fig:gamma}
\end{figure}

Stochastic Gross-Pitaevskii equations have been benchmarked against numerous Bose gas experiments in various geometries \cite{Weiler08a,liu2018dynamical,cockburn2012ab,Rooney2013,eckel2018rapidly,ota2018collisionless,bland2020persistent}, including Bose-Bose mixtures \cite{de2014quenched}. These comparisons to experimental data include direct modeling of the evaporative cooling process \cite{Weiler08a,liu2018dynamical}. In these works, $\gamma$ is approximated by fitting the condensate atom number growth rate to experimental observations. However, there is also an approximate analytic solution appropriate for near-equilibrium solutions that depends on the chemical potential, energy cut-off, temperature, and interaction strength \cite{Gardiner2003a}. One comparison found $\gamma$ extracted from condensate growth data is an order of magnitude larger than the analytic approximation \cite{liu2018dynamical}. The choice of $\gamma$ does not affect the equilibrium properties of the system \cite{rooney2012stochastic} (due to the fluctuation-dissipation theorem \cite{kubo1966fluctuation,Linscott2014a}), however it affects many observables during equilibration.

To the best of our knowledge there does not exist any analytic prediction for $\gamma$ for the dipolar system \cite{Linscott2014a}. We approximate $\gamma$ based on direct experiment-theory comparisons with the condensate growth rate in Ref.~\cite{Sohmen2021}. For that case we have the relevant experimental data available. There, the supersolid formation was studied in detail for a dysprosium supersolid in a cigar-shaped geometry, and a value of $\gamma = 0.0075$ was found to give quantitatively similar growth behavior as observed experimentally. Here, for a qualitatively similar regime \footnote{Despite the different trap geometries, $(f_x,f_y,f_z) = (36,88,141)$Hz in Ref.~\cite{Sohmen2021} compared to $(f_x,f_y,f_z) = (33,33,167)$Hz, the peak densities are similar $n_\text{peak} = 1.6\times10^{21}\mu\text{m}^{-3}$ compared to $n_\text{peak} = 1.8\times10^{21}\mu\text{m}^{-3}$ and thus will have similar quality supersolids.} we initially assume the same value.

In Fig.~\ref{fig:gamma} we investigate the $\gamma$ dependency on the evaporative cooling protocol presented in the main text, namely a $^{164}$Dy gas in a pancake trap with $(f_x,f_y,f_z) = (33,33,167)$Hz and $a_s = 88a_0$. Further details are given in the main text. We consider both half and double the initial value, i.e.~${\gamma = (0.00375, 0.0075, 0.015)}$. Interestingly, several observables are sensitive to the choice of $\gamma$. This includes: atom number versus time, onset time of global phase coherence versus onset time of crystal structure, and the number of free vortices trapped within the crystal. For our simulations we calculate the phase coherence $C^p$ and density overlap $C^d$. Despite the c-field atom number increasing faster with larger $\gamma$, the growth of phase coherence does not appear as clearly dependent on $\gamma$. This is possibly due to the number of vortices generated through the quench, which have been shown to appear more readily with increasing $\gamma$, although there is also evidence they are damped quicker too \cite{Weiler08a}. Curves for simulations with a long-lived single vortex are shown with a lower opacity in Fig.~\ref{fig:gamma}(b), as this greatly influences $C^p$, and these simulations are not included in the averages shown in the main paper. It is also worth noting that we do not see free vortices after the interaction quenches. Even in the presence of a free vortex the final state can still be considered as a coherent supersolid, with $C^d\sim1$, however $C^p<1$. Future improvements to this measure could involve finding the vortex centre and multiplying the phase by the opposite circulation. This is not easy however due to the nonlinear azimuthal phase profile of a vortex in a supersolid \cite{ancilotto2021vortex,roccuzzo2020rotating}.

\begin{figure}[!t]
\begin{centering}
    \includegraphics[width=0.45\textwidth]{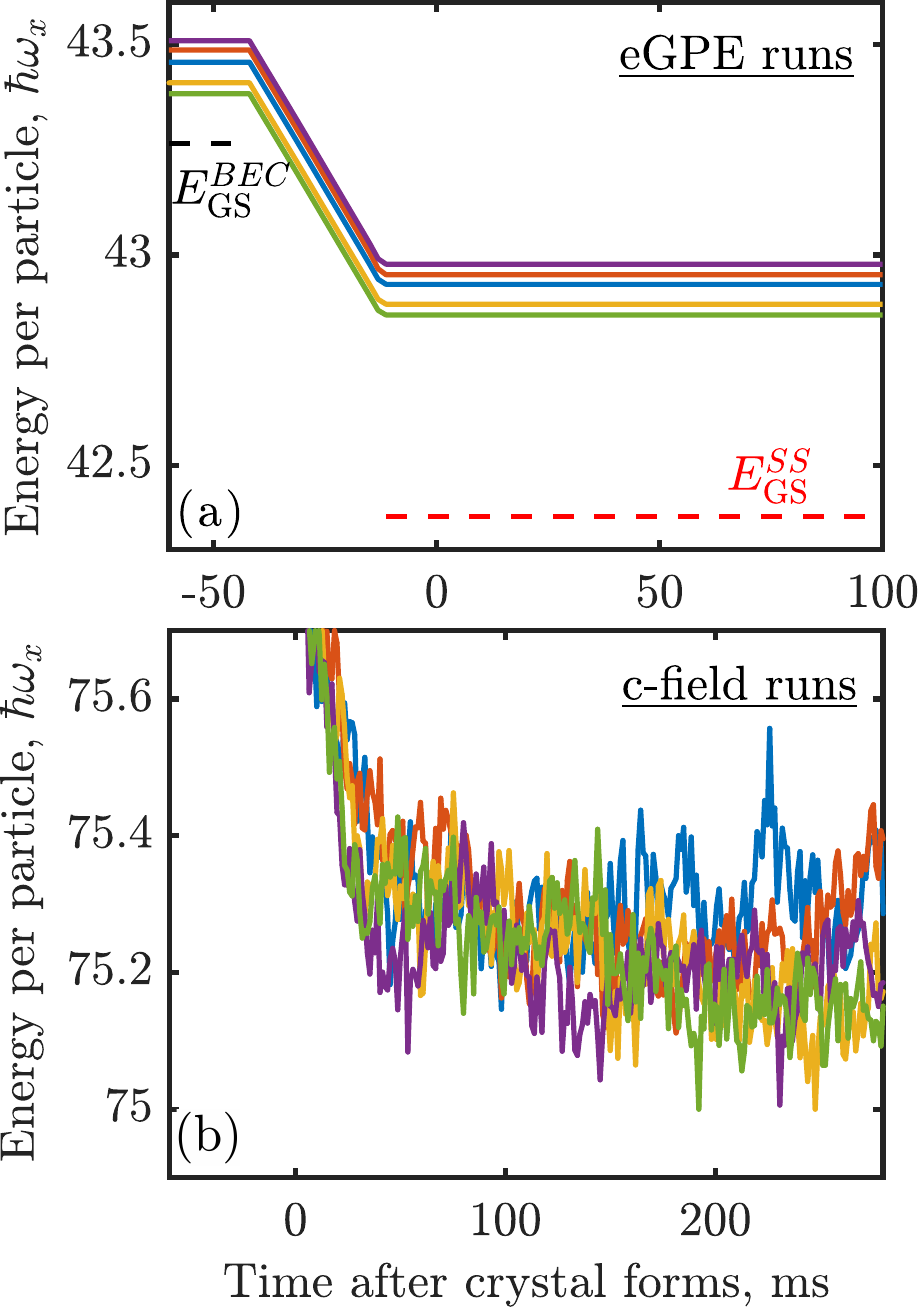}
\caption{Energy versus time for the pancake quench simulation runs considered in Fig.~2 of the main text. (a) Interaction quenches simulated using the eGPE with $N\approx 2.1\times10^5$. Dashed lines show the ground state energy for the initial parameters, $E_{\rm GS}^{\rm BEC}$, and the final parameters corresponding to the target supersolid, $ E_{\rm GS}^{\rm SS}$. (b) Temperature quenches simulated with the SeGPE. All curves are with fixed $\gamma=0.0075$. For both subplots $f_{x,y,z}= (33,33,167)$\,Hz.
}
\label{Fig:GPEenergy}
\end{centering}
\end{figure}

\begin{figure*}[!t]
\begin{centering}
    \includegraphics[width=0.9\textwidth]{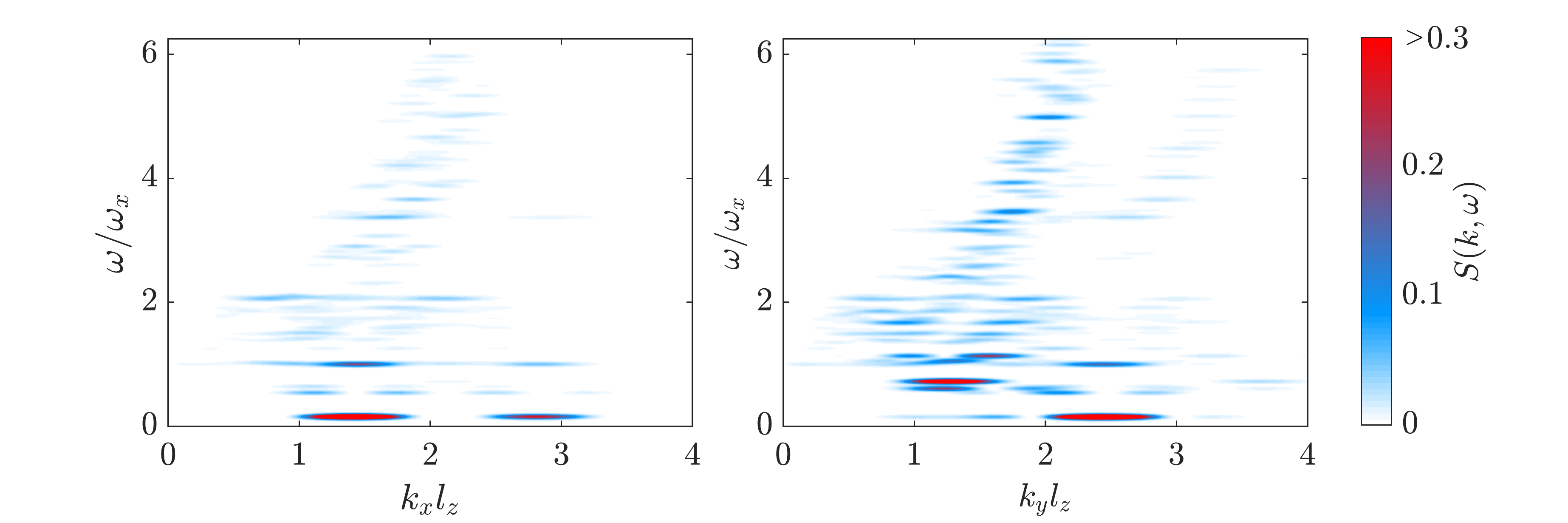}
\caption{Dynamic structure factor (DSF), Eq.~\eqref{Eq:SF}, normalized to peak value, in energy-momentum space for a $^{164}$Dy 7-droplet hexagon supersolid. Left: DSF along $k_x$ with $k_y=k_z=0$. Right: DSF along $k_y$ with $k_x=k_z=0$. Parameters: $a_s=90a_0$, $f_{x,y,z}=(52.83,52.83,167)$\,Hz, $N=9.5\times10^4$.}
\label{Fig:DSF}
\end{centering}
\end{figure*}

The effect of varying $\gamma$ is most obvious in the overlap between the simulation density and ground state density, $C^d$ [Fig.~\ref{fig:gamma}(c)]. Larger $\gamma$ forces the c-field atom number, and hence density, to rapidly increase, forcing the fast production of droplets. We believe that comparing a spectrum of observables such as these will provide important benchmarks to fine tune the simulations, and that this will also provide an important test in the future for the development of analytic theories that can predict $\gamma$. As supersolid production in 2D becomes more routine, direct comparison between the condensate atom number and droplet number growth rates in particular will become crucial in determining the appropriate choice of $\gamma$.

It is worth noting that even if the supersolid formation time was a few 100ms longer than the data presented here, the whole process would still be faster than evaporatively cooling into the BEC state, quenching the interactions and then waiting for the phase coherence to reappear. In this latter scenario, significant three-body losses play a negative role. Previous works in 1D have maximized phase coherence by increasing the final $a_s$, and hence increasing the superfluid connection between droplets \cite{petter2020high}, and decreasing the droplet peak density. However, the droplet number is strongly dependent on $a_s$, and we find that such a strategy significantly decreases the number of droplets.

\section{The role of energy during supersolid formation}
It is instructive to investigate the role of energy during the formation of 2D droplet arrays, via both interaction quenches and the temperature quenches. In Fig.~\ref{Fig:GPEenergy} (a), we show the energy versus time for the five interaction quench simulations considered in Fig.~1(a) of the main text. We have also marked the energy of the ground state before the interaction quench ($E_{\rm GS}^{\rm BEC}$), and the energy of the ground state following the interaction quench ($E_{\rm GS}^{\rm SS}$), with the superscript indicating that the ground state is initially in the BEC phase, then later in the supersolid phase. An estimate for the energy added by crossing the phase transition can be evaluated as $[E(t_{\rm final})-E_{\rm GS}^{\rm SS}] - [E(t_{\rm initial})-E_{\rm GS}^{\rm BEC}] \approx 0.35 \hbar\omega_x$. Note that $E(t_{\rm initial})-E_{\rm GS}^{\rm BEC}$ does not equal zero due to the random noise added to the initial state.

It is interesting at this point to compare the final energies of the eGPE simulations in Fig.~\ref{Fig:GPEenergy} (a) with the final energies of the c-field simulations following the evaporative cooling quench, which are shown in Fig.~\ref{Fig:GPEenergy} (b), which differ by more than $30\hbar\omega_x$. From such a comparison we can deduce that the important factor for disrupting supersolid formation following an interacting quench is not so much the total energy injected into the system by crossing the first-order phase transition [in Fig.~\ref{Fig:GPEenergy} (a)] but, rather, which modes become excited. This large disparity in energy, however, tells us that at much longer time scales than shown here, the eGPE states may settle down to a better quality supersolid than the SeGPE [see e.g. Fig.~2 of the main text], but this could be on the order of seconds, much larger than the supersolid life time.

From our dynamic structure factor calculations shown in Fig.~\ref{Fig:DSF} (which will be discussed shortly), one can see that the out-of-phase Goldstone modes (the low-energy modes at finite momentum that show up as red ovals) are particularly vulnerable to excitation by the interaction quench. Note that although this figure is for the 7-droplet supersolid rather than the 19-droplet one, such low-energy Goldstone modes are a general feature of dipolar supersolids \cite{natale2019excitation,guo2019low}. Since these modes inherently cause both phase and crystal excitations, they directly act to disrupt the supersolid. Furthermore, we also see some vortex pairs after the interaction quench, and these also play a role. Interestingly, even in the interaction-quench simulations, a supersolid is obtained in the long-time limit (although too long to be useful for current experiments due to lifetime limitations), even though the total energy is conserved, thanks to a damping of these phase and crystal excitations.

\section{Supersolid excitations}

\begin{figure}
    \centering
    \includegraphics[width=3.38in]{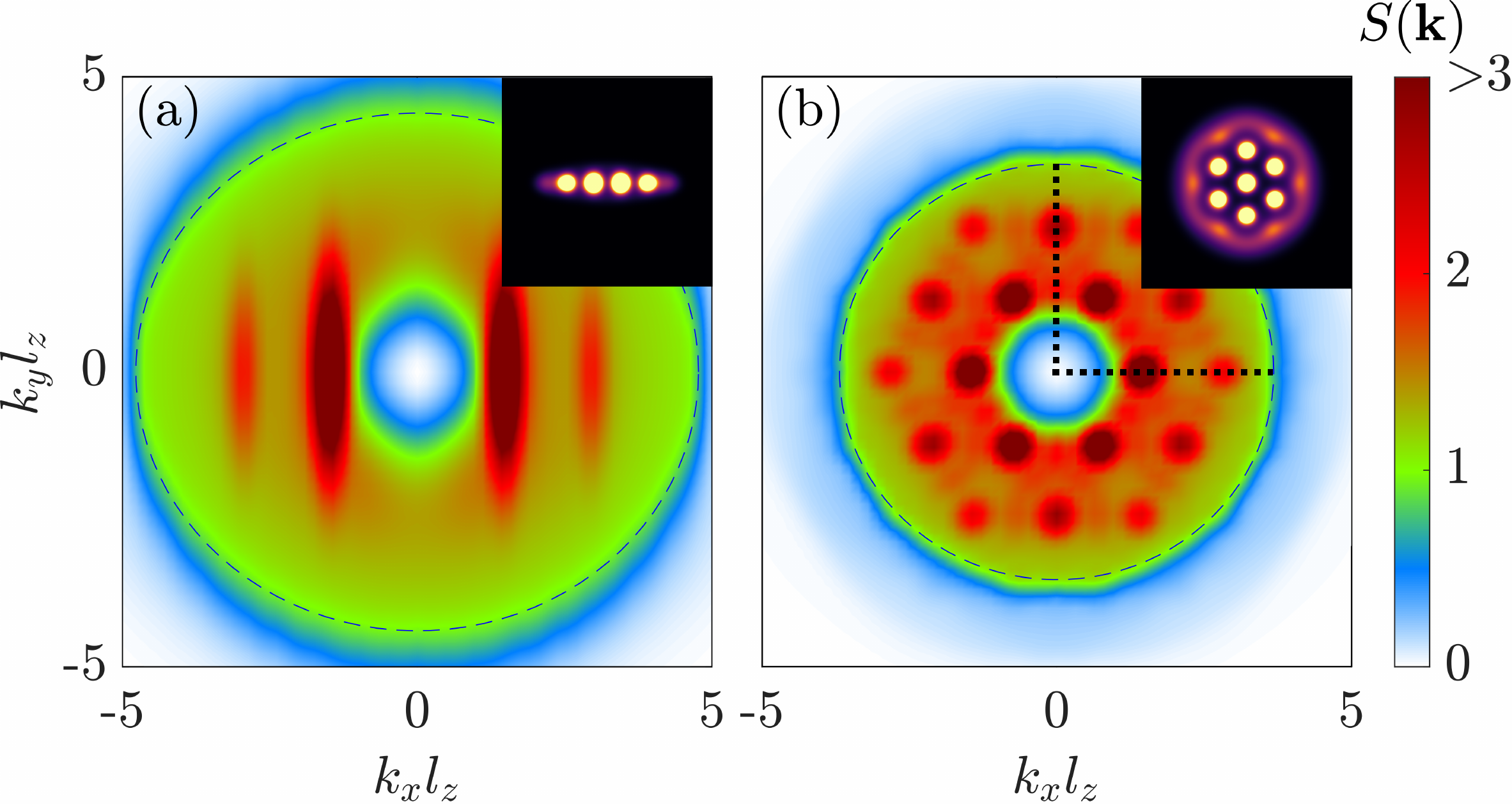}
\caption {Static structure factors for (a) linear and (b) 7-droplet hexagon supersolids.
Convergence is achieved within the dashed ellipses (see text). Dotted lines in (b) correspond to the trajectories shown in Fig.~\ref{Fig:DSF}, integrated over energy.
Parameters for linear chain: $a_s=90a_0$, $f_{x,y,z}=(52.83,130,167)$\,Hz, $N=4\times10^4$. For 2D crystal: $a_s=90a_0$, $f_{x,y,z}=(52.83,52.83,167)$\,Hz, $N=9.5\times10^4$.}
	 \label{fig:4} 
\end{figure}

We investigate the 7-droplet hexagon supersolid, the same configuration as shown in Fig.~3 of the main text, using the extended Gross-Pitaevskii equation (eGPE), focusing here on its excitations. We perform a Bogoliubov-de Gennes linearization and present results in the form of the dynamic structure factor,
\begin{align}
S(\textbf{k},\omega) = \sum_l\left|\int \dx[u_l^*(\textbf{r})+v_l^*(\textbf{x})]e^{i\textbf{k}\cdot\textbf{x}}\psi_0(\textbf{x})\right|^2\delta(\omega-\omega_l)\,, \label{Eq:SF}
\end{align}
Here, $\psi_0$ is the ground state wavefunction normalized to unity, i.e.~$\int \text{d}^3\mathbf{x}|\psi_0(\mathbf{x})|^2=1$, and $\left \{u_l(\textbf{x}),v_l(\textbf{x})\right\}$ are the quasiparticle excitations with energy $\omega_l$ \cite{Zambelli2000,blakie2002theory}.
The dynamic structure factor along two orthogonal directions is displayed in Fig.~\ref{Fig:DSF}. Note, the asymmetry along $k_x$ and $k_y$ is due to the triangular configuration of the crystal, which can be seen clearly in Fig.~\ref{fig:4}(b).

To explore the role of dimensionality, Fig.~\ref{fig:4} compares the static structure factor, $S(\textbf{k}) = \int \text{d}\omega\, S(\textbf{k},\omega)$, for both linear and 7-droplet hexagon supersolids.
These results are converged within the dashed ellipses, set by ensuring that the $f$-sum rule, $\int \text{d}\omega\,\omega S(k,\omega)=\hbar^2k^2/2m$, is satisfied, and should be ignored outside. Convergence is limited by the number of BdG modes, for which we use 512 modes for both cases.
In order to make a fair comparison between a 1D and 2D supersolids we choose to approximately match the average 2D trap density by fixing $\varrho = Nf_xf_y$ \cite{poli2021maintaining}. As previously reported, the structure factor for the linear case [Fig.~\ref{fig:4}(a)] has peaks corresponding to the average inter-droplet spacing ($2.67\mu$m), $k_xl_z\approx1.43$, and subsequent peaks at integer multiples of this \cite{natale2019excitation}.
We find that the dominant contributing modes to the structure factor peaks are low energy out-of-phase Goldstone modes \cite{goldstone1961field,guo2019low,natale2019excitation}, where the superfluid current and crystal oscillate out-of-phase with one another. Note that for possible comparison with experiments our spectrum in Fig.~\ref{fig:4} was energy broadened with a Gaussian of width $\sigma = 0.008 \,\hbar \omega_z$, note that Fig.~1(b) of the main text was similarly broadened by $\sigma = 0.004 \,\hbar \omega_z$.

For the 2D supersolid, Fig.~\ref{fig:4}(b) displays peaks situated at $k_\rho l_z\approx1.43$ every 60$^\circ$ azimuthally, where ${k_\rho = \sqrt{k_x^2+k_y^2}}$.
These peaks reflect the hexagonal structure of the ground state, however this value does {\it not} directly reflect the inter-droplet spacing ($3.05\mu$m, which would correspond to $k_\rho l_z\approx1.25$), but rather the spacing of lattice planes between droplets. Crucially, the six inner momentum peaks are rotated compared to the droplet crystal, analogous to what we observed experimentally in the TOF images.
Similar to the 1D chain, we find that the out-of-phase Goldstone modes--a manifestation of superfluidity--contribute to the majority of the peak signal.

\end{document}